\documentclass[aps,showpacs,twocolumn]{revtex4-1}
\usepackage[english]{babel}
\usepackage{graphicx}
\usepackage{stmaryrd}
\usepackage{amssymb}
\usepackage{amsfonts}
\usepackage{amsmath}
\usepackage{mathrsfs}
\usepackage{braket}
\usepackage{color}
\usepackage{xspace}

\usepackage{bm}
\usepackage{hyperref}

\begin{document}

\title{Spontaneously magnetized Tomonaga-Luttinger liquid in frustrated
quantum antiferromagnets}
\author{Shunsuke C. Furuya}
\author{Thierry Giamarchi}
\affiliation
{DPMC-MaNEP, University of Geneva, 24 Quai Ernest-Ansermet
CH-1211 Geneva, Switzerland
}
\date{\today}
\begin{abstract}
We develop a theory of spontaneously magnetized Tomonaga-Luttinger
liquid (SMTLL) in geometrically frustrated quasi-one-dimensional quantum
magnets by taking an $S=1/2$ ferrimagnet on a union-jack lattice as
an example.
We show that a strong frustration leads to a spontaneous
magnetization because of the ferrimagnetic nature of lattice.
Due to the ferrimagnetic order, the local magnetization has an
incommensurate oscillation with the position.
We show that the spontaneously magnetized TLL is smoothly connected
to the existence of a Nambu-Goldstone boson in the canted ferrimagnetic phase of
a two-dimensional frustrated antiferromagnet.
\end{abstract}
\maketitle

\section{Introduction}\label{sec:introduction}

Spontaneous symmetry breaking is one of the most fundamental
concepts in physics.
It provides the mechanism to generate mass of elementary
particles and allows macroscopic alignment of magnetic moments in
ferromagnets.
Spontaneous breaking of global continuous symmetry is accompanied by a
massless  excitation, the Nambu-Goldstone
boson~\cite{Nambu_1961,Goldstone_1961,Goldstone_1962}. 
Since Nambu-Goldstone boson governs low-energy physics at long distance
and the low-energy physics is affected by the geometry of
system, the dimensionality of the system has strong influences on
Nambu-Goldstone boson. 

Such effects are most prominent in one dimension (1D)
because of the large suppression of ordering at finite
temperatures~\cite{MerminWagner_1966,Hohenberg_1967}
and even at zero temperature~\cite{Momoi_AF_1996} due to quantum effects.
As a result the breaking of a continuous symmetry in 1D is deemed impossible.
For systems such as a 1D superfluid, indeed no long range order exists,
and the proper description is the one of a Tomonaga-Luttinger
liquid (TLL)~\cite{Giamarchi_book}. Despite the absence of the true
long-range order~\cite{Cazalilla_1D_2011,DelMaestro_SF_2011}, Goldstone
modes exist, and have a dynamical origin~\cite{Eggel_SF_2011}.
However in some rare cases, such as a ferromagnet, the ground state can, even
in 1D spontaneously break a continuous symmetry. This prompts immediately for
the question of why and for which systems such phenomena can occur.

In order to shed light on the possibility of
spontaneous symmetry breaking in 1D, dynamical aspects of
the system need to be carefully considered. In addition, in view of the recent
experimental progresses in realizing 1D quantum liquids in various
situations~\cite{Stroferle_BEC_2004,Taniguchi_SF_2010,Klanjsek_BPCB_2008,Jeong_DIMPY_2013},
it is worthwhile to search for novel manifestations of spontaneous
symmetry breaking in 1D. 

For quantum magnetism in 1D, one expects a system with quasi-long range
antiferromagnetic order to have a relativistic dispersion, which is the
case of the TLL, while a ferromagnet would have a quadratic one. This
behavior of the Nambu-Goldstone boson has been formulated in a quite general 
context~\cite{Nielsen_NG_1976,Watanabe_NG_2012,Hidaka_NG_2013}. Finding
in 1D a system that would spontaneously break the continuous
rotational symmetry of the spins, while at the same time retaining some
TLL behavior would thus be interesting and an example of the more
general character of Nambu-Goldstone boson. 

A very good possibility to realize such a spontaneously magnetized TLL
(SMTLL) is offered by ferrimagnetic systems. Several numerical
studies
\cite{Ivanov_ferri_1998,Yoshikawa_ferri_2005,Hida_ferri_2007,
hida_ferri_2008,Montenegro-Filho_ferri_2008,Hida_ferri_2010,Shimokawa_UJ_2012} 
have followed this route and found incommensurate ferrimagnetic phases
which can be candidates for the SMTLL. However besides the numerical
results, there is still no microscopic theory that explains the nature
of the incommensurate phase and could relate it to a SMTLL. 

\begin{figure}[b!]
 \centering
 \includegraphics[bb= 0 0 900 350, width=\linewidth]{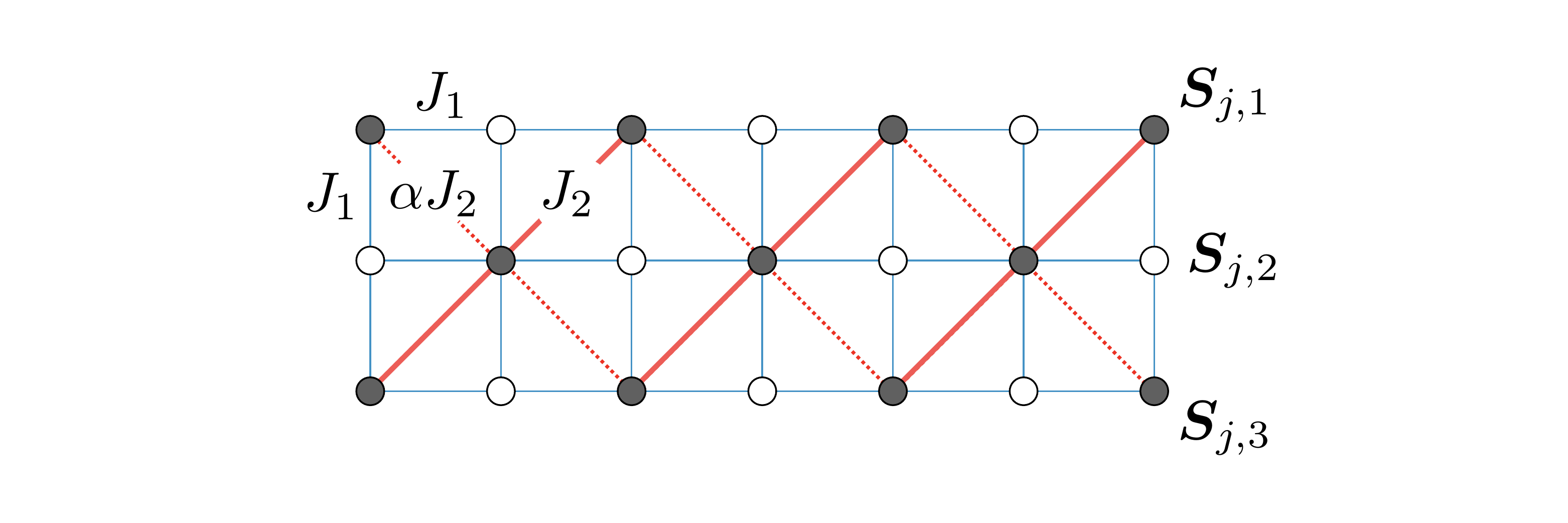}
 \caption{
 The three-leg union-jack ladder \eqref{eq:H_lat}.
 }
 \label{fig:lattice}
\end{figure}

In this paper, we present such a theory of the SMTLL,
showing that one can have simultaneously a spontaneous breaking of the
spin-rotation
symmetry leading to a finite magnetization and a TLL behavior.
We demonstrate that this SMTLL phase is realized
in the ground state of an
$S=1/2$ geometrically frustrated quantum antiferromagnet on a 1D array
of the union jack lattice [see Fig.~\ref{fig:lattice}].

\section{Instability of the Tomonaga-Luttinger liquid}\label{sec:inst}

We consider the union jack (UJ) spin Hamiltonian:
\begin{align}
 \mathcal H
 & = J_1 \sum_j \sum_{a=1}^3 \bm S_{j,a} \cdot \bm S_{j+1,a} + J_1
 \sum_{j} \bm S_{j,2} \cdot (\bm S_{j,1} + \bm S_{j,3}) \notag \\
 &  +  J_2 \sum_j \bm S_{2j,2}\cdot (\bm S_{2j-1,3} + \bm
 S_{2j+1,1}) \notag \\
 & + \alpha J_2 \sum_j \bm S_{2j,2}\cdot (\bm S_{2j-1,1} +  \bm
 S_{2j+1,3}),
 \label{eq:H_lat}
\end{align}
where $J_{1,2}>0$ and $0<\alpha\ll1$.
The parameter $\alpha$ denotes the imbalance of the diagonal
interactions. Throughout the Paper, we fix $\alpha$
and change the ratio $J_2/J_1$ from $0$ to $+\infty$. Note that this
model has \emph{a priori} full spin-rotational symmetry. 

\subsection{Classical ground state}

We first consider the classical ground state minimizing the
energy of a unit cell.
The classical analysis on the UJ ladder \eqref{eq:H_lat}
is similar to the 2D UJ
antiferromagnet~\cite{Collins_UJ_2006,Zheng_UJ_2007,Bishop_UJ_2010}.
For $0\le J_2/J_1<1/2$, the classical ground state is the N\'eel state.
For $1/2<J_2/J_1$, spins on the filled sites in Fig.~\ref{fig:lattice}~(a)
become canted with a polar angle $\vartheta=\cos^{-1}(J_1/2J_2)$ and
the classical ground state in the canted phase has an incommensurate
magnetization,
\[
 \braket{S^z_{j,a}}=\frac{\hbar S}{2}\biggl(1-\frac{J_1}{2J_2}\biggr).
\]
Hereafter we use $\hbar=1$ for simplicity.
At the classical level, a spontaneous magnetization occurs.
However, since quantum fluctuation usually destroys long-range order
in 1D systems \eqref{eq:H_lat}, we have to take them into account
to conclude on the existence of a spontaneous magnetization in 1D.

\subsection{The Tomonaga-Luttinger liquid}

\begin{figure}[b!]
 \centering
 \includegraphics[bb= 0 0 900 300, width=\linewidth]{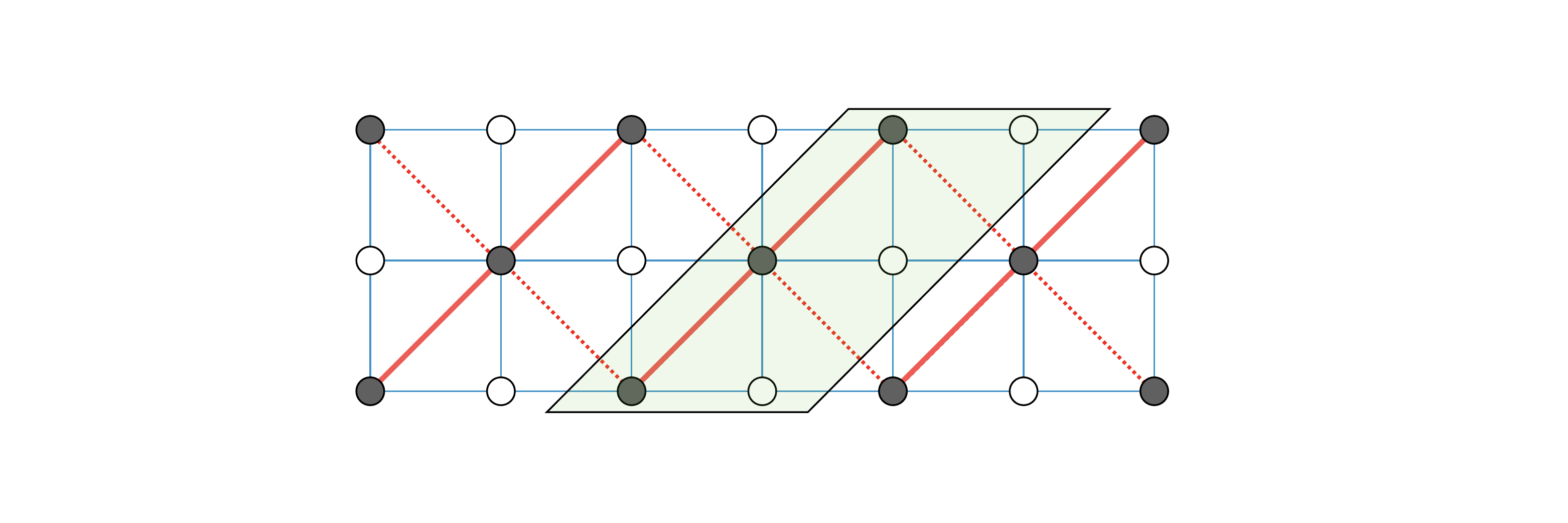}
 \caption{
 The shaded area depicts the unit cell of the $O(3)$ nonlinear $\sigma$
 model. 
 }
 \label{fig:diagonal}
\end{figure}

To do so we derive the low-energy effective field theory of the UJ ladder
\eqref{eq:H_lat}.
When the diagonal interaction is small enough, $J_2/J_1\ll1$,
the low-energy effective field theory is written as a function of two
slowly varying fields
\begin{align}
 \bm{n}&=\frac 1{2S}\sum_{a=1}^3(\bm{S}_{2j+2-a,a}-\bm{S}_{2j+3-a,a}),
 \label{eq:n_def}
 \\
\bm{l}&=\frac 1{2a_0}\sum_{a=1}^3(\bm{S}_{2j+2-a,a}+\bm{S}_{2j+3-a,a}). 
 \label{eq:l_def}
\end{align}
Here $a_0$ is the lattice spacing.
We take a diagonal unit cell (Fig.~\ref{fig:diagonal}) to define
$\bm{n}$ and $\bm{l}$ along the $J_2$ bond~\cite{Sierra_review_1996}.
The $\bm{n}$ and $\bm{l}$ fields denote respectively staggered and
uniform magnetization densities and satisfy the constraints
\begin{equation}
 f(\bm{n},\bm{l})\equiv\bm{n}^2-1-\frac{1}{S}-\frac{\bm{l}^2}{S^2}=0 
  \label{eq:const}
\end{equation}
and $\bm{n}\cdot\bm{l}=0$.
The constraint \eqref{eq:const} is usually replaceable to $\bm{n}^2=1$.
However, the $\bm{l}^2$ term will play an essential role for our purpose.

The Hamiltonian \eqref{eq:H_lat} in the low-energy limit is given
by
\begin{align}
 \mathcal{H}
 &= \int \frac{dx}2 \biggl[ \frac 1{\sum_{a,b}\mathcal M^{-1}_{a,b}} \bm
 l^2+ 2S^2\sum_a p_a (\partial_x\bm n)^2 \notag \\
 & \quad
 +\frac{2S\sum_{a,b} p_a\mathcal M^{-1}_{a,b}}{\sum_{a,b}\mathcal
 M^{-1}_{a,b}} (\bm l \cdot \partial_x\bm n + \partial_x\bm n \cdot \bm
 l)\biggr]  
 \label{eq:H_NLSM_nl_expanded} \\
 &=  \int dx  \biggl[
  \frac{gv}2\biggl(\bm{l}
  -\frac{\Theta}{4\pi}\partial_x\bm{n}\biggr)^2
  +\frac{v}{2g}(\partial_x\bm{n})^2
  \biggr]
  ,
  \label{eq:H_NLSM_nl}
\end{align}
where $g$ is a coupling constant, $v$ is the velocity, $\Theta$ is a
topological angle given by
\begin{align}
 g
 &=\frac 1S \biggl[2\sum_{a,b,c}p_a\mathcal{M}^{-1}_{b,c}
 -\biggl(\frac{\Theta}{4\pi S}\biggr)^2\biggr]^{-1/2},
 \label{eq:g} \\
 v
 &=Sa_0\biggl[\frac{2\sum_ap_a}{\sum_{b,c}\mathcal{M}^{-1}_{b,c}}
 -\biggl(\frac{\Theta}{4\pi S}\frac
 1{\sum_{a,b}\mathcal{M}^{-1}_{a,b}}\biggr)^2\biggr]^{1/2}, 
 \label{eq:v} \\
 \Theta &=6\pi S,
 \label{eq:Theta}
\end{align}
and $p_a=3J_1/2+(J_1/2)\delta_{a,2}$.
While $g$ and $v$ depend on a $3\times3$ matrix of microscopic
parameters,
\begin{equation}
 \mathcal{M}=
  \begin{pmatrix}
   5J_1-\tilde{J}_2 & J_1-\tilde{J}_2 & 0 \\
   J_1-\tilde{J}_2 & 6J_1-2\tilde{J}_2 & J_1-\tilde{J}_2 \\
   0 & J_1-\tilde{J}_2 & 5J_1-\tilde{J}_2
  \end{pmatrix}
  \label{eq:matrix}
\end{equation}
with $\tilde{J}_2=(1+\alpha)J_2/2$,
the topological angle \eqref{eq:Theta} is
determined only by the number of legs.
The derivation of the effective field theory~\eqref{eq:H_NLSM_nl} is
explained in the case of the three-leg spin ladder in
Refs.~\onlinecite{Sierra_NLSM_1996,Sierra_review_1996}. 
We obtain the effective field theory \eqref{eq:H_NLSM_nl} by replacing
the rung coupling of the three-leg ladder to $J_1-\tilde J_2$ 
in the matrix \eqref{eq:matrix}. 
We can see from Eq.~\eqref{eq:H_NLSM_nl_expanded} that
information on the structure of the UJ lattice
\eqref{eq:H_lat} is encoded in the matrix \eqref{eq:matrix}.
Integrating $\bm{l}$ out, we obtain~\cite{Affleck_review_1989}
\begin{align}
 \mathcal{H}
 &= \frac{v}{2g}\int dx
  \biggl[\frac 1{v^2}(\partial_\tau
  \bm{n})^2+(\partial_x\bm{n})^2\biggr] +i\Theta Q,
  \label{eq:H_NLSM}
\end{align}
where $\bm{n}^2=1$ and 
\[
 Q=\frac 1{4\pi}\int_{-\infty}^\infty dx\, \bm{n}\cdot
 \partial_\tau\bm{n} \times\partial_x\bm{n}
\]
gives an integer after integrating with an imaginary time $\tau$, that
is, $\int_{-\infty}^\infty d\tau \, Q \in\mathbb Z$.
The Hamiltonian \eqref{eq:H_NLSM} is the $O(3)$ nonlinear
$\sigma$ model (NLSM). 
The $\Theta$ term controls the low-energy limit of the NLSM
\eqref{eq:H_NLSM}~\cite{Haldane_NLSM_1983}.
When $\Theta\equiv\pi$ (mod $2\pi$), namely when
$S$ is a half integer,
the $O(3)$ NLSM \eqref{eq:H_NLSM} is identical to
the TLL as a conformal field theory with a central charge
$c=1$~\cite{Affleck_spinS_1987},
\begin{equation}
 \mathcal H = \frac v{2\pi}\int dx\biggl[K(\partial_x\theta)^2+\frac
  1K(\partial_x\phi)^2\biggr].
  \label{eq:H_TLL}
\end{equation}
We use here the notation for the TLL of Ref.~\onlinecite{Giamarchi_book}.
The nature of the $\phi$ and $\theta$ fields will be discussed in detail
later. Since $\Theta$ is independent of $J_2$, for small enough $J_2$,
the diagonal interaction is irrelevant and
the UJ ladder \eqref{eq:H_lat} has the TLL ground state \eqref{eq:H_TLL}, and thus
in particular zero spontaneous magnetization. 

\subsection{Instability at $k=0$}

However the diagonal interaction has a serious impact on the
ground state, and lead to an instability of the TLL.
The diagonal interaction $J_2$ partly compensates $J_1$ in the
matrix \eqref{eq:matrix} and
it reduces the velocity down to $v=0$ where the
linearization of the dispersion relation $\omega=vk$ becomes invalid.
Let us denote the instability point as $J_2^{c1}$.
The instability point is determined from the zeros of the matrix
\eqref{eq:matrix}.
The matrix $\mathcal M$ is positive definite for $\tilde J_2=0$.
As we increase $\tilde J_2$, the positive definiteness first breaks down
at $\tilde J_2=7 J_1/3$, namely
\begin{equation}
 J_2^{c1} = \frac{14 J_1}{3(1+\alpha)}.
  \label{eq:J_2^c1}
\end{equation}
When $0<J_2^{c1}-J_2\ll J_1$, $g$ and $v$ can be expanded with respect to
$(J_2^{c1}-J_2)/J_1$.
In fact, since $\sum_{a,b}\mathcal M^{-1}_{a,b}\sim
[(J_2-J_2^{c1})/J_1]^{-1}$,  we obtain
$g\propto S^{-1}[(J_2^{c1}-J_2)/J_1]^{1/2}$ and
\begin{equation}
 v\propto J_1S\biggl(\frac{J_2^{c1}-J_2}{J_1}\biggr)^{1/2}.
  \label{eq:v_c1}
\end{equation}
We can easily see that the velocity \eqref{eq:v_c1}
approaches zero when $J_2\nearrow J_2^{c1}$.
Note that the TLL parameter $K$ must be $1/2$ to ensure that the
Hamiltonian \eqref{eq:H_TLL} preserves the $SU(2)$ rotational
symmetry~\cite{Giamarchi_book}.
Thus, the susceptibility $\chi=K/\pi v$ of the TLL
\eqref{eq:H_TLL} diverges as $J_2\nearrow J_2^{c1}$.

Near the instability point $J_2=J_2^{c1}$, a careful treatment
of the interaction is required.
The interaction of the $O(3)$ NLSM
\eqref{eq:H_NLSM} is non-perturbatively included in the constraint
\eqref{eq:const}.
Namely, the partition function $Z$ of the $O(3)$ NLSM \eqref{eq:H_NLSM}
is given as a path integral,
\begin{equation}
 Z = \int \mathcal D \bm n \mathcal D\bm l \delta\bigl(f(\bm n, \bm
  l)\bigr) \delta(\bm n \cdot \bm l)\exp\biggl(-\int d\tau \mathcal
  H\biggr).
  \label{eq:Z}
\end{equation}
The constraint \eqref{eq:const} generates a strong repulsion
$\lambda(\bm{l}^2)^2$. 
Indeed, one can add a strongly repulsive interaction
$U\{f(\bm{n},\bm{l})\}^2$ with $U\nearrow+\infty$ to the
Hamiltonian \eqref{eq:H_NLSM_nl} instead of imposing the constraint
\eqref{eq:const}.
Then, the partition function \eqref{eq:Z} is approximated as
\begin{align}
 Z
 &=\int \mathcal D \bm n \mathcal D\bm l \delta(\bm n\cdot \bm l)
 \notag
 \\
 &\quad \times
  \exp\biggl(- U\int d\tau dx \, \bigl\{f(\bm
 n\cdot \bm l)\bigr\}^2\biggr)\exp\biggl(-\int d\tau \mathcal
 H\biggr)\notag \\ 
 &=\int \mathcal D \bm n \mathcal D\bm l \delta(\bm n\cdot \bm l) 
 \exp\biggl(-\int d\tau \bar{\mathcal H}\biggr),
\end{align}
where we obtain the Hamiltonian $\bar{\mathcal H}$
with the effective repulsion,
\begin{align}
 \bar{\mathcal{H}}
 &=\int dx
 \biggl[\frac{gv}2\bm{L}^2+ \lambda(\bm{L}^2)^2
  +\frac{v}{2g}(\partial_x\bm{n})^2 +U(\bm{n}^2-1)^2
  \biggr],
  \label{eq:H_NLSM_L4}
\end{align}
where $\lambda=U/S^4>0$  and
\begin{equation}
 \bm{L}\equiv\bm{l}-\frac{\Theta}{4\pi}\partial_x\bm{n}.
  \label{eq:L_def}
\end{equation}
We introduced the quartic interaction $(\bm{L}^2)^2$ instead of
$(\bm{l}^2)^2$ because their difference
(e.g. $\{(\partial_x\bm{n})^2\}^2$) is negligible as we see below.

When $J_2>J_2^{c1}$, the following inequalities are valid:
\begin{align}
 gv&= \frac 1{\sum_{a,b}\mathcal M^{-1}_{a,b}}<0,
 \label{eq:ineq_gv} \\
 \frac vg &=2S^2\sum_a p_a
 -\biggl(\frac{\Theta}{4\pi}\biggr)^2\frac 1{\sum_{a,b}\mathcal
 M^{-1}_{a,b}}>0. 
 \label{eq:ineq_v/g}
\end{align}
Thus, the interaction that the $\bm L$ field feels takes a form of the
wine bottle:
\begin{equation}
 \frac{gv}2\bm{L}^2+\lambda(\bm{L}^2)^2
  =\lambda\biggl(\bm{L}^2-\frac{1}{\lambda|gv|}\biggr)^2+\mathrm{const.}
  \label{eq:bottle}
\end{equation}
The potential \eqref{eq:bottle} leads to a nonvanishing expectation
value of $\bm{L}$. 
Note that the instability occurs only in the uniform part of the 
$O(3)$ NLSM \eqref{eq:H_NLSM_L4} because of the inequalities
\eqref{eq:ineq_gv} and \eqref{eq:ineq_v/g}.
The diagonal coupling only changes the sign of the coupling constant 
of the $\bm{L}^2$ [Eq.~\eqref{eq:ineq_gv}] keeping
that of $(\partial_x\bm{n})^2$ positive (Eq.~\eqref{eq:ineq_v/g}).
Namely, the instability only occurs at the wave number $k=0$ and the
nonzero expectation value of the $\bm L$ field 
\eqref{eq:L_def} is attributed to the magnetization density $\bm{l}$.
Therefore the ground state has a \emph{spontaneous magnetization} per site,
$\bm{M}\equiv\langle\bm{l}\rangle/3$, with 
\begin{equation}
 |\bm{M}|=\frac{1}{3}\biggl(\frac 1{\lambda|gv|}\biggr)^{1/2} \propto
  \biggl(\frac{J_2-J_2^{c1}}{J_1}\biggr)^{1/2}.
  \label{eq:mag}
\end{equation}
The transition around $J_2^{c1}$ is described by a Ginzburg-Landau like theory of
second order transitions.
Contrarily to the $(\bm{L}^2)^2$ term,
the higher-order interaction $\{(\partial_x\bm{n})^2\}^2$ is negligible
because the lower-order term $(\partial_x\bm{n})^2$ is stable.
Since $\lambda>0$ the transition cannot be first order.
We emphasize that our derivation of the spontaneous magnetization
\eqref{eq:mag} fully respects the $SU(2)$ rotational symmetry and $\bm{M}$
can point in an arbitrary direction.

\subsection{Nambu-Goldstone bosons}

Let us explain the nature of Nambu-Goldstone boson generated from this phase transition.
First we focus on the $k=0$ part.
We rewrite the Hamiltonian in terms of fluctuation
\begin{equation}
  \bm{m}\equiv\bm{L}-3\bm{M}.
   \label{eq:m_def}
\end{equation}
We can assume
$\bm{M}=M(0\>0\>1)^T$ without loss of generality.
The longitudinal component $m^z$ has a mass $\Delta=12\lambda M^2$
because of the interaction \eqref{eq:bottle}, that is,
$\lambda(\bm{L}^2-9M^2)^2\simeq12\lambda M^2(m^z)^2$.
The transverse component
$\bm{m}^\perp\equiv(m^x,m^y,0)$ is
the Nambu-Goldstone boson generated from
the spontaneous magnetization and it possesses a nonrelativistic
dispersion relation~\cite{Brauner_NG_2010,Brauner_NG_2007}.
Dispersion relations of the longitudinal and the transverse modes are
respectively
\begin{equation}
 E_\parallel(k)=\Delta+\frac{k^2}{2m_\parallel}, \quad
  E_\perp(k)=\frac{k^2}{2m_\perp}.
  \label{eq:E_k=0}
\end{equation}

We can find another massless excitation near $k=\pi$.
Then the Hamiltonian \eqref{eq:H_NLSM_L4} turns into
\begin{equation}
 \bar{\mathcal H}
  = \int dx \biggl[4\lambda M^2\bm{m}^2 +\frac v{2g} (\partial_x\bm{n})^2
  \biggr] +\mathcal H',
  \label{eq:H_SMTLL_nm}
\end{equation}
The repulsion $U(\bm{n}^2-1)^2$
is transformed into the constraint $\bm{n}^2=1$ again.
The last term of Eq.~\eqref{eq:H_SMTLL_nm} denotes the anisotropy that
the spontaneous magnetization induces,
\begin{equation}
 \mathcal H' = V\int dx \, \bigl\{
  2(m^z)^2-(m^x)^2-(m^y)^2\bigr\},
  \label{eq:aniso}
\end{equation}
with $V=4\lambda M^2$, which seemingly equals to the coupling constant of
$\bm{m}^2$. 
However, after including renormalization due to irrelevant 
operators, the coupling constant $V$ of the anisotropic interaction
\eqref{eq:aniso} actually deviates from that of the isotropic part
$\bm{m}^2$. 
Here we first omit the anisotropy \eqref{eq:aniso} and include it later
perturbatively because it does not modify qualitative features of the 
effective Hamiltonian of the Nambu-Goldstone boson.
Let us rewrite Hamiltonian \eqref{eq:H_SMTLL_nm} in terms of $\bm n$ and
$\bm l$,
\begin{equation}
 \bar{\mathcal H} = \int dx \biggl[\frac{\bar gu}2\biggl(\bm
  l-\frac{\Theta}{4\pi} \partial_x\bm n\biggr)^2 + \frac
  u{2\bar g}(\partial_x\bm n)^2 -3\bar gu\bm M \cdot \bm l\biggr].
  \label{eq:H_SMTLL_nl}
\end{equation}
Here we introduced the coupling constant $\bar g$ and the velocity $u$
in a parallel manner as Eq.~\eqref{eq:H_NLSM_nl}.
They are given by
\begin{align}
 \bar{g}&=
 \frac{2M}S \biggl[2\lambda \sum_a p_a
 -\biggl(\frac{\Theta}{4\pi S}\biggr)^2 \frac{\lambda}{\sum_{a,b}\mathcal
 M^{-1}_{a,b}}\biggr]^{-1/2}
 \label{eq:barg} \\
 u&=2MSa_0\biggl[2\lambda \sum_a p_a -\biggl(\frac{\Theta}{4\pi
 S}\biggr)^2 \frac{\lambda}{\sum_{a,b}\mathcal M^{-1}_{a,b}}\biggr]^{1/2}.
 \label{eq:u}
\end{align}
Note that both $\bar g$ and $u$ are positive
and proportional to the
magnitude of the spontaneous magnetization $|\bm M|$.

The last term of the Hamiltonian \eqref{eq:H_SMTLL_nl} can be seen as
the Zeeman energy $-\bm{h}_{\rm eff}\cdot \bm S_{j,l}$.
For further understanding of the effective Hamiltonian
\eqref{eq:H_SMTLL_nl} of the Nambu-Goldstone boson at $k=\pi$,
we integrate $\bm l$ out:
\begin{align}
 \bar{\mathcal{H}}
 &= \frac{u}{2\bar{g}}\int dx \biggl[\frac
 1{u^2}(\partial_\tau \bm{n}+i\bm{h}_{\mathrm{eff}}\times
 \bm{n})^2  +(\partial_x\bm{n})^2\biggr] \notag \\
 & \quad + 9\bar{g}u \int dx \, (\bm{M}\cdot\bm{n})^2 +i\Theta Q,
 \label{eq:H_NLSM_mag}
\end{align}
where $\bm h_{\rm eff}$ is written as
\begin{equation}
  \bm{h}_{\mathrm{eff}}=3\bar{g}u\bm{M}.
\end{equation}
If we include the perturbation $\mathcal H'$ at lowest order,
it gives a correction
\begin{equation}
\mathcal H'\simeq V\int dx (\bm M \cdot \bm n)^2 [3(n^z)^2-1] 
\end{equation}
to the Hamiltonian \eqref{eq:H_NLSM_mag}.
If we use the value $V=4\lambda M$, $\mathcal H'$ replaces the term
$9\bar gu(\bm M\cdot \bm n)^2=9\bar gu M^2(n^z)^2$ of
Eq.~\eqref{eq:H_NLSM_mag} with $27\bar gu M^2(n^z)^4$, which has no
impact on qualitative aspects of the effective field theory
\eqref{eq:H_NLSM_mag}.

The $O(3)$ NLSM \eqref{eq:H_NLSM_mag} leads to three important
consequences.
First, the spontaneous magnetization $\bm{M}$ leaves
$\Theta\equiv\pi$ (mod $2\pi$) intact.
Second, $\bm{M}$ generates the easy-plane anisotropy.
Finally, the $O(3)$ NLSM \eqref{eq:H_NLSM_mag} is semiclassical.
While the NLSM \eqref{eq:H_NLSM} in the TLL phase
has a coupling $g\propto1/S$, the
NLSM \eqref{eq:H_NLSM_mag} has $\bar{g}\propto M/S$ [see
Eq.~\eqref{eq:barg}]. 
Thus, the NLSM \eqref{eq:H_NLSM_mag} behaves similarly to a
spin-$S_{\mathrm{eff}}$ Heisenberg antiferromagnetic chain with
a large half-integer spin $S_{\mathrm{eff}}\sim S/M$.

The effective field theory
at $k=\pi$ is thus, the TLL under an effective magnetic field
$\bm{h}_{\mathrm{eff}}=h_{\mathrm{eff}}(0\>0\>1)^T$; that is,
\begin{equation}
 \bar{\mathcal{H}} = \frac{u}{2\pi}\int dx\,
  \biggl[\bar{K}(\partial_x\theta)^2
  +\frac1{\bar{K}}(\partial_x\phi)^2\biggr]
  -\frac{h_{\mathrm{eff}}}{\pi}\int dx \,\partial_x\phi.
  \label{eq:H_SMTLL}
\end{equation}
which is the effective model for the SMTLL.
The TLL parameter $\bar{K}$ is determined from the
relation~\cite{Giamarchi_book} 
\begin{equation}
 M=\frac{h_{\mathrm{eff}}\bar{K}}{\pi u}. 
  \label{eq:M2K}
\end{equation}
The TLL parameter
\begin{equation}
 \bar{K}=\frac{\pi}{\bar{g}}\propto
  \biggl(\frac{J_2-J_2^{c1}}{J_1}\biggr)^{-1/2} 
\end{equation}
diverges at the instability point $J_2^{c1}$.
The point $J_2^{c1}$ brings about a divergence of the
susceptibility, 
\begin{equation}
 \chi\propto \biggl(\frac{|J_2^{c1}-J_2|}{J_1}\biggr)^{-\gamma},
\end{equation}
with the critical exponent $\gamma$ is given by
$\gamma=1/2$ for $J_2\nearrow J_2^{c1}$ and
$\gamma=1$ for $J_2\searrow J_2^{c1}$.

\subsection{Dynamical structure factors}

We now use this theory to compute the dynamical
structure factors in the SMTLL phase.
We focus on longitudinal and transverse dynamical structure factors,
$S^\parallel(k,\omega)=\int_{-\infty}^\infty dtdx\,e^{i(\omega
t-kx)}\braket{S^z_j(t)S^z_0(0)}$ and
$S^\perp(k,\omega)=\int_{-\infty}^\infty dtdx\,e^{i(\omega
t-kx)}\braket{S^+_j(t)S^-_0(0)}$.

\begin{figure}[t!]
 \centering
 \includegraphics[bb= 0 0 1040 520, width=\linewidth]{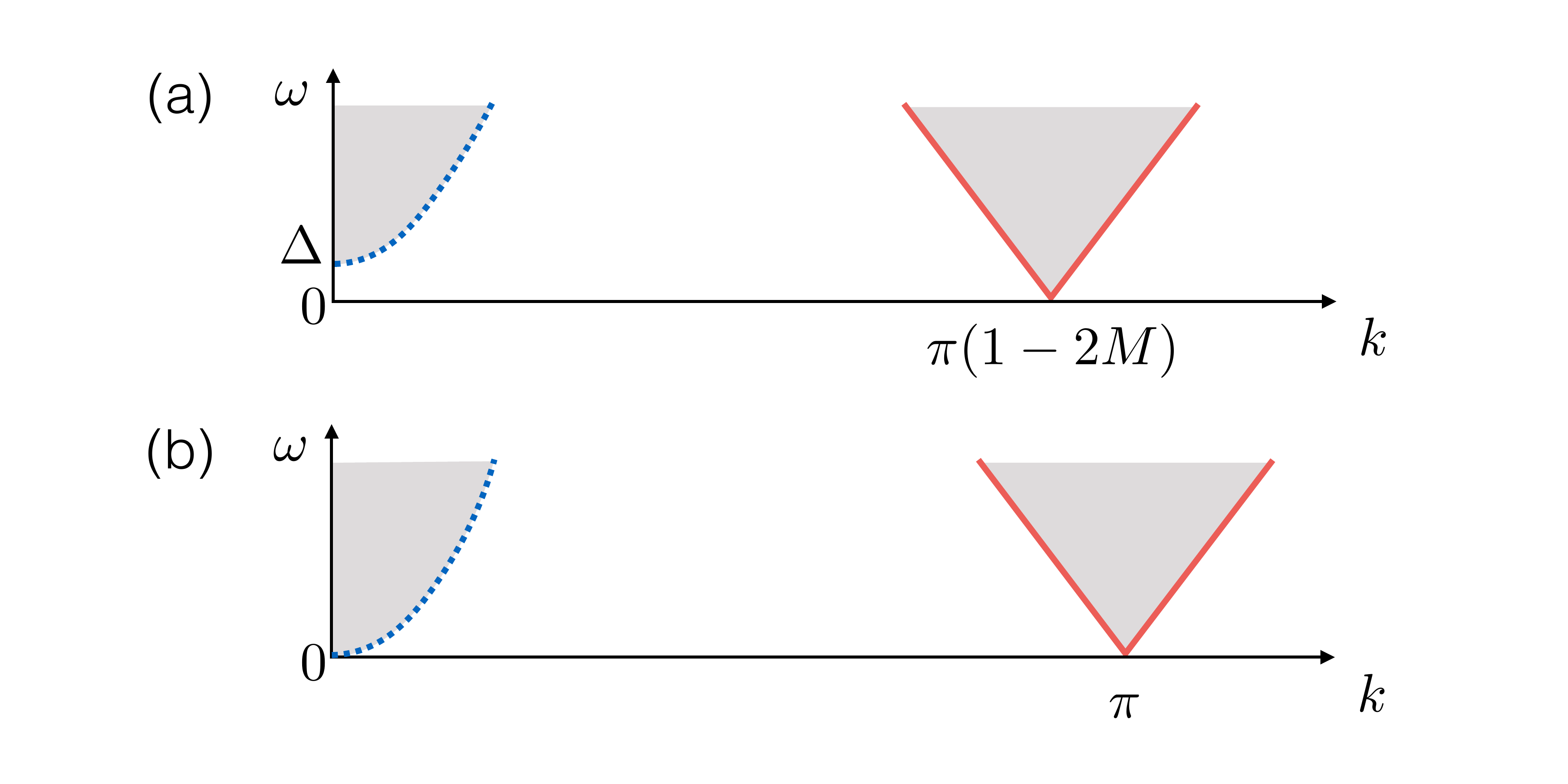}
 \caption{
 Dynamical structure factors (a) $S^\parallel(k,\omega)$ and (b)
 $S^\perp(k,\omega)$ in the low-energy region.
 Outside of the shaded area, the dynamical structure factor has zero
 intensity. 
 The red lines represent the linear dispersion of the SMTLL and
 the blue dashed curves represent the quadratic dispersions
 \eqref{eq:E_k=0} of the
 nonrelativistic Goldstone mode [$E_\perp(k)$] and the massive
 mode [$E_\parallel(k)$].
 }
 \label{fig:DSF}
\end{figure}

The longitudinal and transverse dynamical structure factors near $k=\pi$
are the same as 
those of the $S=1/2$ Heisenberg antiferromagnetic chain under
magnetic field~\cite{Chitra_1997,Giamarchi_book},
\begin{align}
 &S^\parallel(k=\pi(1-2M)+\delta k, \omega) \notag \\
 &=
 \frac{\pi^2C_\parallel}{u\Gamma^2(\bar{K})}\theta_H(\omega-u|\delta
 k|)\biggl(\frac{4u^2}{\omega^2-u^2(\delta k)^2}\biggr)^{1-\bar K},
 \label{eq:Szz_pi} \\
 &S^\perp(k=\pi+\delta k, \omega) \notag \\
 &= \frac{\pi^2C_\perp}{u\Gamma^2(\frac
 1{4\bar K})}\theta_H(\omega-u|\delta
 k|)\biggl(\frac{4u^2}{\omega^2-u^2(\delta
 k)^2}\biggr)^{1- 1/4\bar K}.
 \label{eq:Spm_pi}
\end{align}
$\theta_H(z)$ is the Heaviside's step function and $C_\parallel$ and
$C_\perp$ are
nonuniversal constants.
Equations \eqref{eq:Szz_pi} and \eqref{eq:Spm_pi} hold when $|\delta
k|\ll 1$.
The dynamical structure factor near $k=0$ is given by
\begin{equation}
 S^\nu(k=\delta k,\omega)
 =\theta_H\bigl(\omega-E_\nu(\delta k)\bigr)
\frac{C'_\nu}{\omega-E_\nu(\delta k)}, \quad (\nu=\parallel, \perp),
 \label{eq:DSF_0}
\end{equation}
where $C'_\parallel$ and $C'_\perp$ are constants.
Figure~\ref{fig:DSF} shows the longitudinal and transverse dynamical
structure factors in the low-energy region.
The dynamical structure factor near $k=0$ \eqref{eq:DSF_0} 
clearly shows difference between the SMTLL and either a TLL under magnetic
field (e.g. the $S=1/2$ antiferromagnetic chain~\cite{Giamarchi_book})
or a field-induced
TLL (e.g. the $S=1/2$ two-leg spin ladder~\cite{Bouillot_ladder_2011}), 
for which the symmetry has been externally broken. 

\section{Commensurate phase}\label{sec:comm}

\subsection{Commensurability condition}

Let us now examine the behavior upon increasing $J_2$ further.
The spontaneous magnetization saturates at a certain point
$J_2^{c2}(>J_2^{c1})$.
In the case of the UJ ladder, we can find a saturation condition
in the spirit of
the Oshikawa-Yamanaka-Affleck theory~\cite{Oshikawa_plateau_1997}.
To do so, we need to clarify the physical meanings of the $\phi$ and
$\theta$ fields of  the SMTLL \eqref{eq:H_SMTLL}.
The definitions \eqref{eq:n_def} and \eqref{eq:l_def} of $\bm{n}$ and
$\bm{l}$ indicate that $\bm{n}$ and $\bm{l}$, equivalently
$\phi$ and $\theta$, represent a ``center-of-mass'' mode.
When one consider the UJ ladder \eqref{eq:H_lat} as a system of
three spin chains weakly coupled by the rung and the diagonal
interactions~\cite{footnote1},
each spin chain is equivalent to a TLL written in a compactified boson
$\phi_a$ and its dual $\theta_a$ ($a=1,2,3$).
The bosons $\phi$ and $\theta$ represent the center-of-mass mode because
they are given by $\phi=\phi_1+\phi_2+\phi_3$
and $\theta=\theta_1+\theta_2+\theta_3$.
The other ``relative-motion'' modes, $\phi_1-\phi_2$ and
$\phi_1+\phi_2-2\phi_3$, are massive and negligible
in the low-energy effective field theories
\eqref{eq:H_TLL} and \eqref{eq:H_SMTLL}.
The two-site translational symmetry $j\to j+2$ of the UJ ladder
\eqref{eq:H_lat} requires the invariance of the effective field theory
under the translation 
\begin{equation}
 \phi\to\phi+6(S-M)\pi.
\end{equation}
Given an incommensurate magnetization $M$ satisfying
\begin{equation}
 6(S-M)\not\in\mathbb{Z},
  \label{eq:OYA}
\end{equation}
the incommensurability condition \eqref{eq:OYA} prohibits
relevant interactions of $\phi$, for instance $\cos(2\phi)$, to
appear in the effective field theory \eqref{eq:H_SMTLL}.
Relevant interactions of $\theta$ are not allowed from another reason,
that is, the $U(1)$ symmetry of the ground
state~\cite{Oshikawa_plateau_1997}.

Equation~\eqref{eq:OYA} shows that
the $\phi$ field can be massive when the incommensurability
condition \eqref{eq:OYA} is violated.
Increasing $M$ from zero,
the condition \eqref{eq:OYA} first breaks down when
\begin{equation}
 \frac M{M_s} = \frac 13.
  \label{eq:plateau}
\end{equation}
Here $M_s=S$ is the saturated value of $M$.
Thus, a commensurate phase as the $1/3$-plateau \eqref{eq:plateau}
should exist.
The commensurate phase has only one massless Nambu-Goldstone boson near $k=0$ because the
SMTLL acquires a mass from a relevant interaction $\cos(2\phi)$.
The condition \eqref{eq:plateau} gives the saturation condition of the
UJ ladder.

\subsection{Trimer-spin chain}

In order to complete the above derivation of the spontaneous
magnetization, we show that for large $J_2$ it can be shown to
occur directly from the lattice model \eqref{eq:H_lat}.

\begin{figure}[b!]
 \centering
 \includegraphics[bb = 0 0 1200 500,
 width=\linewidth]{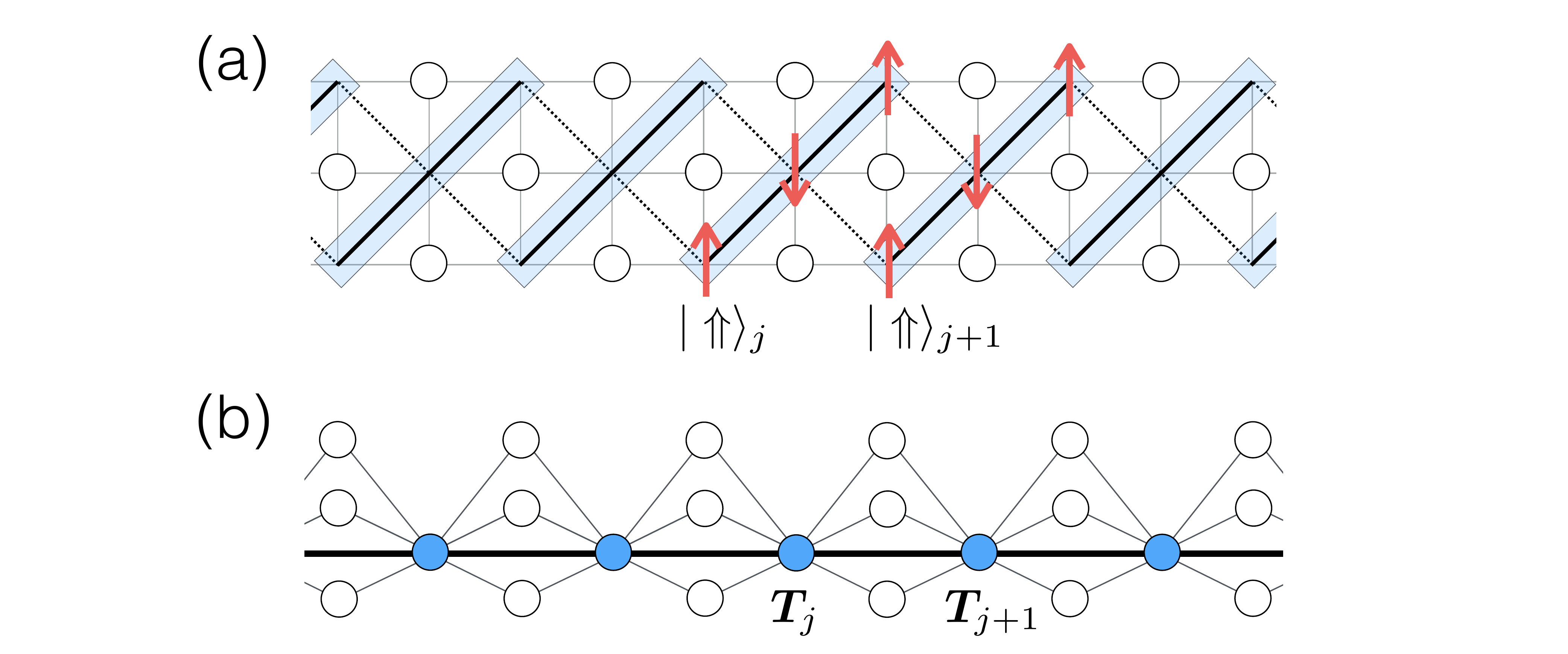}
 \caption{
 (a) A configuration of trimers (solid rectangles) at $J_2/J_1\gg1$.
 Trimers are surrounded by spins (circles).
 In the low-energy limit, one can regard the trimer as the $S=1/2$
 pseudo spin.
 The trimer-trimer interaction is ferromagnetic and the
 trimer-spin interaction is antiferromagnetic.
 (b) The effective model that describes the commensurate phase.
 The solid and blank circles represent trimers ($S=1/2$ pseudo spins)
 and $S=1/2$  spins, respectively. 
 The thick line represents the ferromagnetic coupling \eqref{eq:J_F}
 of trimers 
 and the thin lines are the antiferromagnetic coupling of trimers and
 spins.
 }
 \label{fig:trimer}
\end{figure}

The commensurate phase is identified as a ferromagnetic phase of trimers
formed on diagonal $J_2$ bonds (Fig.~\ref{fig:trimer}~(a)).
Let us consider the case $J_2/J_1\gg1$.
When $J_1=0$, the UJ ladder \eqref{eq:H_lat} is composed of an $S=1/2$
diamond chain~\cite{Takano_diamond_1996,Tonegawa_diamond_2000}
and isolated spins.
Three spins $\bm{S}_{2j+1,1},\bm{S}_{2j,2}$ and
$\bm{S}_{2j-1,3}$  form a trimer [a solid rectangle in
Fig.~\ref{fig:lattice}~(b)] on the strongest $J_2$ bond.
To describe the ground state and the lowest-energy excitation, we may
replace the three spins with an $S=1/2$ pseudo
spin~\cite{Tonegawa_diamond_2000}, 
\begin{equation}
  \bm{S}_{2j+1,1}=\bm{S}_{2j-1,3}=\frac23\bm{T}_j, \quad
   \bm{S}_{2j,2}=-\frac13\bm{T}_j.
   \label{eq:S2T}
\end{equation}
The eigenstates, $|\Uparrow\rangle_j$ with $T_j^z=1/2$
and $|\Downarrow\rangle_j$ with $T_j^z=-1/2$ are written as
\begin{align}
 |\Uparrow\rangle_j
 &= \frac 1{\sqrt 6}
 (
 |\downarrow\rangle_{j,1}|\uparrow\rangle_{j,2}|\uparrow\rangle_{j,3} 
 -2|\uparrow\rangle_{j,1}|\downarrow\rangle_{j,2}|\uparrow\rangle_{j,3} 
 \notag \\
 & \quad
 +|\uparrow\rangle_{j,1}|\uparrow\rangle_{j,2}|\downarrow\rangle_{j,3} 
 ), \\
 |\Downarrow\rangle_j
 &= \frac 1{\sqrt 6}
 (
 |\uparrow\rangle_{j,1}|\downarrow\rangle_{j,2}|\downarrow\rangle_{j,3} 
 -2|\downarrow\rangle_{j,1}|\uparrow\rangle_{j,2}|\downarrow\rangle_{j,3}
 \notag \\
 & \quad
 +|\downarrow\rangle_{j,1}|\downarrow\rangle_{j,2}|\uparrow\rangle_{j,3} 
 ).
\end{align}
Here $|\uparrow\rangle_{j,a}$ and $|\downarrow\rangle_{j,a}$ are the
eigenstate of $S^z_{2j+2-a,a}$  ($a=1,2,3$).
Figure~\ref{fig:trimer}~(a) depicts that
the mapping from spins to a trimer metamorphoses
an antiferromagnetic interaction $\alpha
J_2\bm{S}_{2j,2}\cdot(\bm{S}_{2j-1,1}+\bm{S}_{2j+1,3})$
to a \emph{ferromagnetic}
trimer-trimer interaction,
\begin{equation}
\alpha J_2\bm{S}_{2j,2}\cdot(\bm{S}_{2j-1,1}+\bm{S}_{2j+1,3})
 =-J_{\mathrm{F}}\bm{T}_j\cdot\bm{T}_{j+1}
 \label{eq:J_F}
\end{equation}
with $J_{\mathrm{F}}=\frac{4\alpha
J_2}9+\mathcal{O}(\alpha^2J_2)$~\cite{Tonegawa_diamond_2000}.
Since the trimer-trimer interaction is ferromagnetic, the
ground state at $J_1=0$ has a nonzero magnetization.

\begin{figure}[b!]
 \centering
 \includegraphics[bb= 0 0 960 320, width=\linewidth]{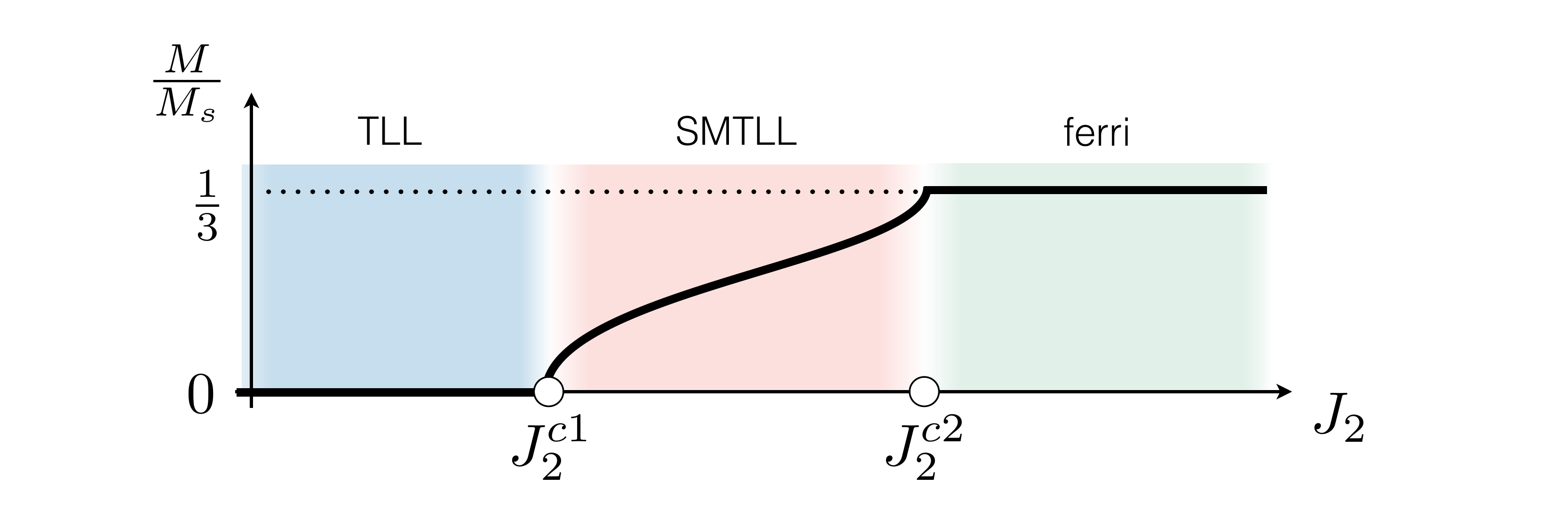}
 \caption{
 A schematic magnetization curve of the UJ ladder \eqref{eq:H_lat}.
 $J_2^{c1}$ and $J_2^{c2}$ represent quantum critical points.
 The SMTLL phase exists in the region $J_2^{c1}<J_2<J_2^{c2}$.
 }
 \label{fig:phase}
\end{figure}

At $J_1=0$, the residual spins [depicted as blank circles in
Fig.~\ref{fig:trimer}~(a) and (b)] are isolated from the trimers.
The nonzero $J_1$ switches on trimer-spin interactions.
The low-energy effective Hamiltonian for $J_2/J_1\gg1$ is given
by
\begin{align}
 \mathcal{H}_{\mathrm{ferri}}
  &= -J_{\mathrm{F}}\sum_{j\in\mathbb{Z}}\bm{T}_j\cdot\bm{T}_{j+1}
 + \sum_{i,j\in\mathbb{Z}} J_{ij}\tilde{\bm{S}}_i\cdot \bm{T}_j,
 \label{eq:H_ferri}
\end{align}
where spins not participating in forming trimers are relabeled as
$\tilde{\bm{S}}_j$.
Fig.~\ref{fig:trimer}~(b) shows interactions of the effective model
\eqref{eq:H_ferri}. 
If we are concerned only with the ground state magnetization of the
trimer-spin chain \eqref{eq:H_ferri},
we do not even need the details of coupling constants.
We use only three facts, $J_{\mathrm{F}}>0$,
$J_{ij}\ge0$ and $\sum_{l\in\mathbb{Z}}J_{il}>0$ for $i,j\in\mathbb{Z}$.
These conditions enable us to apply
the Marshall-Lieb-Mattis
theorem~\cite{Marshall_MLM_1955,Lieb_MLM_1962}
to the model \eqref{eq:H_ferri}.
This theorem imposes that the ground state of the trimer-spin
chain \eqref{eq:H_ferri} must have a fixed magnetization irrespective of
parameters $J_{\mathrm{F}}$ and $J_{ij}$.
The ground state magnetization of the commensurate
phase is exactly Eq.~\eqref{eq:plateau}.
Since the ferromagnetic order of the trimer is exactly the ferrimagnetic
order of spins, the commensurate phase is exactly the
ferrimagnetic phase.
Figure~\ref{fig:phase} shows the ground state magnetization of the UJ ladder,
which reproduces the numerically derived
one~\cite{Shimokawa_UJ_2012} in the $\alpha=1$ case.
The Marshall-Lieb-Mattis theorem allows us even to
take a limit $\alpha\to1$.
However, compared to the numerical result~\cite{Shimokawa_UJ_2012}, we
overestimated $J_2^{c1}$ because of the imbalance $\alpha\ll1$.

\section{Relation to the general theorem}

Now that we have a description of the SMTLL and its Nambu-Goldstone
boson, we can 
compare our result with the general
theorem~\cite{Nielsen_NG_1976,Watanabe_NG_2012,Hidaka_NG_2013} claiming
that the number of broken generators of the symmetry group determines
the number of Nambu-Goldstone boson.
The canted phase generally has a nonrelativistic Nambu-Goldstone boson and a relativistic
Nambu-Goldstone boson~\cite{Watanabe_NG_2012,Roman_canted_2000}, which is
true in the 2D UJ
antiferromagnet~\cite{Zheng_UJ_2007}.
In the 1D case \eqref{eq:H_lat}, the $U(1)$ symmetry is recovered for the
ground state as a result of quantum fluctuations.
In the TLL phase, even the full $SU(2)$ symmetry is recovered.
Therefore, we conclude that the general
theorem~\cite{Watanabe_NG_2012,Hidaka_NG_2013} is applicable to 1D
systems at the classical level.

The SMTLL phase results from the competition of the quasi-long-range
N\'eel order of the TLL and the ferrimagnetic order.
The geometrical frustration is necessary for the SMTLL to exist
because, in the absence of geometrical frustration, the
Marshall-Lieb-Mattis theorem
prohibits the existence of an incommensurate magnetization \eqref{eq:OYA}.
However, frustration alone is not sufficient.
A frustrated diamond chain~\cite{Yoshikawa_ferri_2005} has no incommensurate phase.
This is because the diamond chain cannot have the TLL and the
ferrimagnetic structures simultaneously.
By contrast, the UJ ladder is a superposition of the
three-leg ladder leading to the TLL and the diamond chain leading to the
ferrimagnetic order. In this respect an investigation of
itinerant ferrimagnet~\cite{Ishizuka_ferri_2012} in 1D would be very interesting.

In conclusion, in this Paper, we showed the existence of a spontaneously
magnetized phase, with TLL properties, the SMTLL.
We gave an effective theory for this phase and computed the magnetization
and the dynamical structure factors. We derived the nonrelativistic
Nambu-Goldstone boson near $k=0$ and the SMTLL near $k=\pi$.

\section*{Acknowledgments}

We thank T. Shimokawa for giving us the motivation for this study.
We are also grateful to M. Oshikawa and H. Watanabe for
in-depth discussions.
This work is supported by the Swiss National Foundation under
MaNEP and division II.

\end{document}